\documentclass[preprint]{revtex4}

\usepackage{graphicx}  
\usepackage{bm}        
\usepackage{amssymb}   

\begin{document}

\title{Design of a plasmonic near-field tip for super-resolution IR-imaging}

\author{Fouad Ballout}
\affiliation{Department of Physics, Imperial College London, UK}
\email{f.ballout@imperial.ac.uk}
\author{Erik Br\"{u}ndermann}
\affiliation{Department of Physical Chemistry II, Ruhr-Universit\"{a}t Bochum, Germany}
\email{erik.bruendermann@rub.de}
\author{Diedrich A. Schmidt}
\affiliation{Department of Physical Chemistry II, Ruhr-Universit\"{a}t Bochum, Germany}
\author{Martina Havenith}
\affiliation{Department of Physical Chemistry II, Ruhr-Universit\"{a}t Bochum, Germany}

\begin{abstract}
The concepts of spoof surface plasmon polaritons and adiabatic field compression are employed to design metallic near-field probes 
that allow guiding and focusing of mid-infrared wavelength electromagnetic radiation and provides subwavelength field confinements 
and a lateral optical resolution of about 10 nm.
\end{abstract}
\maketitle

\section{Introduction}
In the past decades scattering scanning near-field microscopy (s-SNOM) in the infrared (IR) spectral region has become an established tool 
for imaging applications including electronic, polymeric and biological nanostructures. After a period of experimental showcases there 
is now a drive to increase its sensitivity and lateral resolution to a level where single molecule detection and imaging becomes accessable. 
However, owing to the long wavelength nature of the involved fields these efforts are limited, unless strategies are pursued
that exploit the internal resonances of the sample and/or tip to enhance the coupling between the probe and the sample. 
Huber et al.~\cite{Huber2007,Huber2008,Huber2010} have utilized phonon polariton resonances of polar crystalline materials to achieve nonlinear 
field enhancements with a lateral resolution of  40 nm at IR and THz wavelengths. As a matter of fact, this approach is 
restricted to materials such as SiC, Si$_3$N$_4$, SiO$_2$ or III-V semiconductors (e.g. GaAs, InP) excluding all materials that do not 
exhibit phononic resonances including biological and soft materials. 

In search of increased flexibility of s-SNOM the alternative approach~\cite{Babadjanyan2000} utilize the shape and internal resonances of the probe 
to excite 
surface plasmon polaritons (SPPs) that can concentrate the electromagnetic radiation into subwavelength regions~\cite{DeAngelis2008,DeAngelis2010}. 
Unfortunately these resonances lie outside the 
chemical fingerprint region which limits the range of applications to optical wavelengths. In order to overcome these limitations, 
Pendry and co-workers~\cite{Pendry2004,Garcia2005} suggested the introduction of a periodic arrangement of 
grooves and holes into the surface of a perfect electric conductor (PEC). This enforces the generation of SPP-like
modes, so-called spoof SPP or designer SPP, at any any desired frequency in dependence of the periodicity. 
So far, this concept of designer SPP has been experimentally demonstrated in the microwave and THz regime
~\cite{Hibbins2005,Williams2008,Zhu2008,Zhao2010}. Out of the number of theoretical investiagtions dealing with spoof SPPs  \cite{Abajo2005,
Chen2006,Maier2006,Maier2006no2,Ruan2007,Dominguez2008,Yang2008,Wang2008,Shen2008,Shen2008no2,Rusina2010,Talebi2010} the work in 
Refs.~\cite{Maier2006,Dominguez2008} is of most interest for the purpose of the work presented here as it utilizes spoof SPPs for THz 
microimaging. The authors proposed a periodically furrowed metallic tapered waveguide structure that focuses THz radiation down to the 
micrometer scale. In this paper we will pick up this strategy and present a novel design of a near-field nanotip which supports surface 
waves in the mid-IR wavelength domain between 5.7 and 6 $\mu$m, containing the dominant lipid and protein absorption bands, and focuses them down
to the 10nm-scale. Our strategy is to find the grating parameters of a corrugated PEC wire to
support spoof SPP modes in the wavelength range 5.7 - 6 $\mu$m. This grating is then grafted onto the surface of a lossy gold tip
onto which a sharp gold nanoantenna is planted. The field enhancement and confinement properties of this hybrid multiscale structure are 
analysed by means of finite-element method (FEM) based electromagnetic simulations.

\begin{figure}
\centering\includegraphics[width=\textwidth]{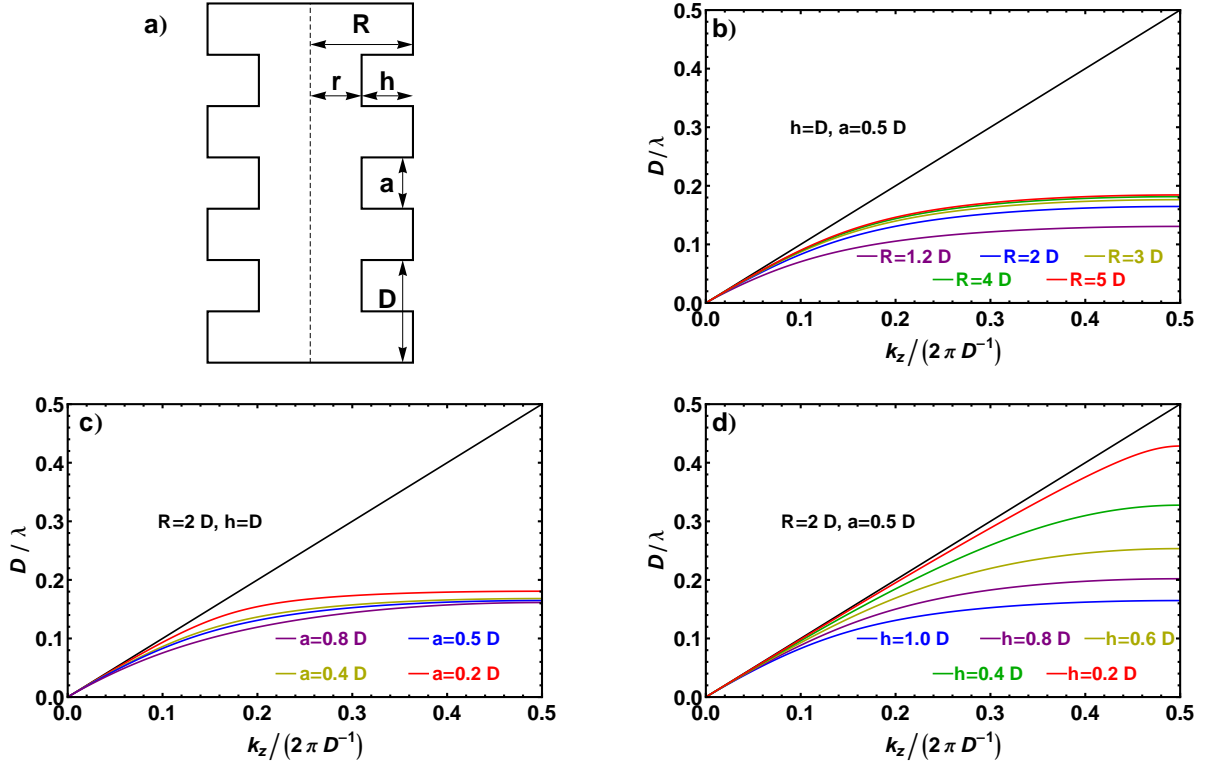}
\caption{a) Scheme of a corrugated wire with grating periodicity \textit{D}, indentation width and depth \textit{a} and \textit{h}, 
respectively. \textit{R} and \textit{r}=\textit{R}-\textit{h} are the outer and inner wire radius, respectively. b)-d) 
Dimensionless
dispersion curves of the bound TM surface wave obtained as solutions to the
implicit Eq.\ref{eq:spoof_SPP_dispersion} and parametrized by b) outer 
wire radius \textit{R}, c) corrugation width \textit{a} and d) indentation depth \textit{h}. The surface wave
propagates vertically along the wire surface in z-direction. For reference the dashed blue curve for \textit{h}=\textit{D}, \textit{a}= 0.5\textit{D}, R= 2\textit{D} is shown in each.}
\label{fig:1}
\end{figure}

\section{Methods}

To find the proper/correct dimensions of corrugations that can confine and guide mid-IR electromagnetic radiation between 5.7 and 
6 $\mu$m along a gold metal tip we will approach the problem as follows:
In the first step, we calculate the dispersion relation of surface waves propagating on an azimuthally corrugated perfect electric 
conducting (i.e. non-dissipative) wire of outer radius \textit{R}, surface grating periodicity \textit{D}, and indentation width and depth \textit{a} and 
\textit{h}, respectively (see. Fig.~\ref{fig:1}a)). To this end, we have to solve numerically the implicit equation 
\cite{Dominguez2008}:
\begin{equation}
\sum_{n=-\infty}^{\infty}\frac{k_0}{q_n}\frac{K_1(Rq_n)}{K_0(Rq_n)}S_n^2
=-\frac{J_1(Rk_0)N_0((R-h)k_0)-J_0((R-h)k_0) N_1(Rk_0)}{J_0(Rk_0)N_0((R-h)k_0)-J_0((R-h)k_0) N_0(Rk_0)},
\label{eq:spoof_SPP_dispersion}
\end{equation}
with $k_0=\omega/c=2\pi\tilde{\nu}$, $q_n=\sqrt{k_n^2-k_0^2}$ and
\begin{equation}
S_n^2=\frac{a}{D}{\rm sinc}^2\left(\frac{k_na}{2}\right),
\end{equation}
where $k_n=k_z+n(2\pi/D)$ with $k_z$ denoting the surface wave vector in z-direction (along the wire axis). $K_m$ is the m-th order 
modified Bessel functions and $J_m$ and $N_m$ are the m-th order Bessel functions of the first and second kind, respectively. 
In order to ensure convergence for the numerical calculation of the SPP dispersion relation, we consider at least four diffraction 
orders in the modal expansion of Eq.~\ref{eq:spoof_SPP_dispersion} (see Ref.~\cite{Shen2008}, Fig.~1). The behavior of the SPP 
dispersion within the first Brillouin zone (i.e. $0\leq k_{z}\leq \pi/D$) in dependence of the geometrical parameters \textit{R}, 
\textit{a} and \textit{h} is depicted in Figs.~\ref{fig:1}b)-d) with all length scales being rescaled by \textit{D}. 
A stronger deviation of the dispersion curve from the light line and thus a stronger confinement of the SPP bands is achieved with 
decreasing \textit{R} and increasing \textit{a} and \textit{h}. 

The specific set of parameters appropiate for our purposes is a
tradeoff between praticality and confinement. On the one hand, we want the width of indentations and protusions to be the same 
to give a nice symmetrical grating pattern, i.e. \textit{a}~=~0.5~\textit{D}. On the other hand we want the spoof SPP band to display a 
strong deviation from the light line indicating a strong confinement of the electromagnetic radiation to the surface, i.e. 
\textit{h}~=~\textit{D}. The corresponding spoof SPP dispersion band exhibits an asymptotic limit at 0.165 = $D/\lambda$. Since we want 
the asymptotic limit of the spoof SPP to appear at 1755 cm$^{-1}$ ($\widehat{=}$ 5698~nm) in order to provide good confinement 
of a surface wave at the lipid and protein absorption wavelengths, \textit{D} has to be 940~nm, which is easily feasible 
with standard ion beam milling techniques. 

\begin{figure}
\centering\includegraphics[width=\textwidth]{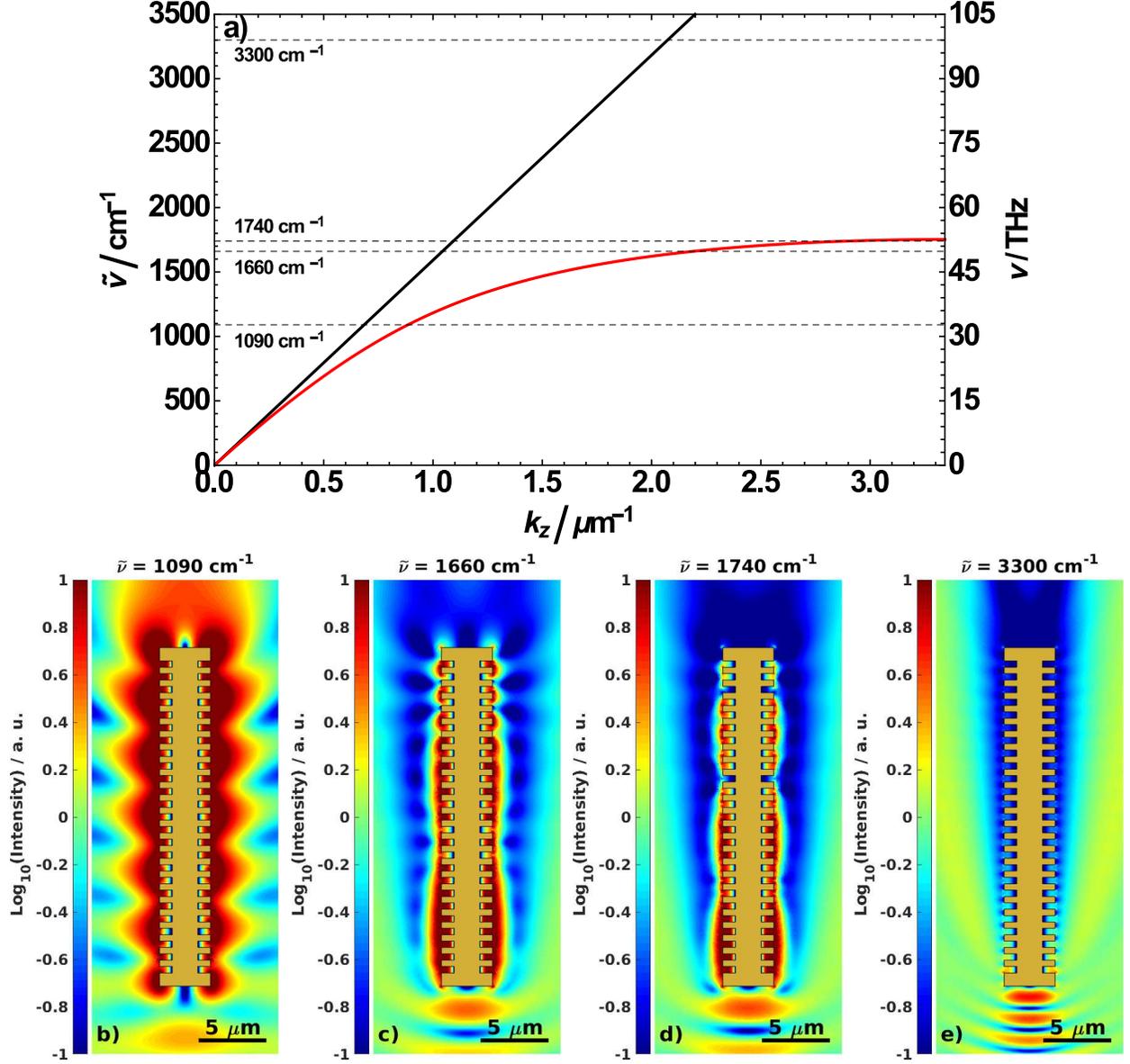}
\caption{a) Dispersion of a spoof SPP propagating along a surface grating with dimensions \textit{D}=940 nm, \textit{a}=470 nm and
\textit{h}=940 nm engraved onto a PEC wire with 3.76 $\mu$m in diameter. b)-e) Simulated electromagnatic field energy 
distribution along the surface of a lossy gold wire of same dimensions illuminated at the bottom by p-polarized light of wavenumbers 
1090, 1660, 1740 and 3300 cm$^{-1}$.} 
\label{fig:2}
\end{figure}

With these specific dimensions for our grating (\textit{D}~=~940~nm, \textit{a}~=~470~nm 
and \textit{h}~=~\textit{D}), we analyse by means of an electromagnetic solver based on the finite element method (JCMSuite by 
JCMWave GmbH) the field distribution around a 25~$\mu$m long, corrugated and lossy gold wire of outer radius 
\textit{R} = 2 \textit{D} = 1.88 $\mu$m when electromagnetic radiation of wavenumber 1090, 1660, 1740 and 3300 cm$^{-1}$ impinges 
in TM-mode (i.e. p-polarized with E-field parallel to the plane of incidence which coincides with the picture plane) from the 
bottom of the wire along its vertical axis. For the frequency dependence of the complex dielectric constant of gold,
we used the Lorent-Drude model with parameters taken from Ref.~\cite{Rakic1998}. According to the the spoof SPP dispersion curve 
belonging to our grating (Fig.~\ref{fig:2}a)), we expect an increasing confinement of the surface waves as we approach the asymptotic 
frequency at 1755~cm$^{-1}$ from the far-infrared end of the spectrum, and no excitation of spoof SPP waves at all in the 
near-infrared frequency range. This grating-induced field/wave dispersion is underpined by the corresponding simulated field 
energy density plots, shown in logarithmic scale in Figs.~\ref{fig:2}b)-e). At 1090 cm$^{-1}$ the surface wave extends far 
outside the region of the wire while they are tightly confined to the wire surface at 1660 cm$^{-1}$ and 1740 cm$^{-1}$. 
At the near-infrared frequency of 3300 cm$^{-1}$, the grating cannot support surface waves and the incident radiation is already 
scattered at the bottom end of the wire. 

\begin{figure}
\centering\includegraphics[width=\textwidth]{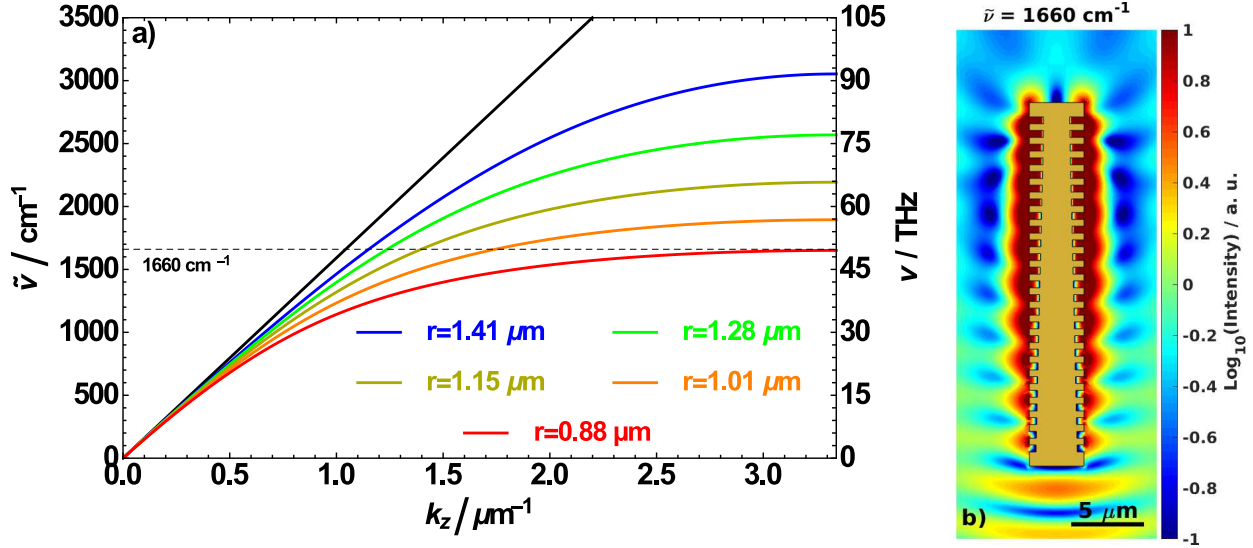}
\caption{a) Spoof SPP dispersion for a set of five PEC wires with different inner radius \textit{r}. 
The residual wire and grating dimensions are kept the same as before. b) Simulated field energy of the spoof SPP 
mode excited at 1660 cm$^{-1}$ and propagating along the surface of a lossy gold wire with the same dimensions as before except 
the inner radius \textit{r} is continuously decreased from 1.41~$\mu$m at the bottom to 0.88~$\mu$m at the top of the wire. This 
illustrates the adiabatic compression of the surface field as it propagates in direction of decreasing inner radius.}
\label{fig:3}
\end{figure}

Since a real s-SNOM experimental setup would not employ a corrugated wire as a probe, we need to replace the metal wire of radius
\textit{R} by a tapered tip that would need to efficiently guide and focus IR radiation as the electromagnetic fields travel along the 
surface grating towards the tip. This concept of adiabatic field compression is illustrated in 
Fig.~\ref{fig:3}a) which shows the spoof SPP dispersion curves for five different corrugated gold wires with the same grating 
parameters \textit{a} and \textit{D} and the same outer radius \textit{R} as before, but different inner radii, 
\textit{r}=\textit{R}-\textit{h}. The spoof SPP band at 1660 cm$^{-1}$ supported on a wire with 1.41~$\mu$m inner radius 
(intersection between the dashed horizontal line and the blue solid line) is quite close to the light line and thus should extend 
over several wavelengths into the surrounding region, while on a wire with an inner radius of 0.88~$\mu$m where the asymptotic 
frequency lies around 1660~cm$^{-1}$, the spoof SPP band (intersection between the dashed horzontal line and the magenta solid line) 
is far from the light line and hence should be tightly confined to the wire surface. This behaviour is also reflected in 
Fig.~\ref{fig:3}b), which shows the simulated field energy density plot for a lossy gold wire at 1660~cm$^{-1}$ with fixed outer radius and variable 
inner radius gradually decreasing from 1.41~$\mu$m at the bottom to 0.88~$\mu$m at the top of the wire. As the surface waves 
travel along the wire from bottom to top the gradual decrease of the inner radius is accompanied by a gradual reduction of the 
group velocity and accordingly an increase in the spatial confinement.

\section{Results and Discussion}

\begin{figure}
\centering\includegraphics[width=\textwidth]{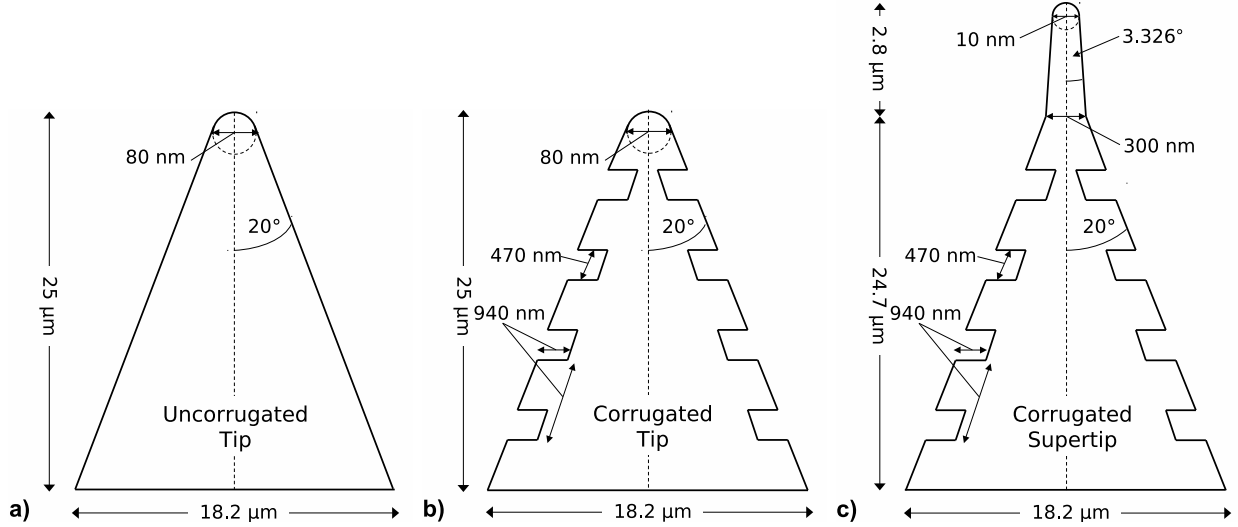}
\caption{Layouts of the near-field probes investigated in this work: a) An uncorrugated tip with dimensions of commercially available 
gold coated scanning probe microscopy tips typical used for IR s-SNOM (e.g. MikroMasch NSC16/Cr–Au). b) A corrugated tip of same 
dimensions as in a) but imprinted with a grating to support spoof SPPs at mid-IR wavelengths (see Fig.~\ref{fig:2}a)). c) A 
corrugated supertip composed of the corrugated tip fitted with a sharp apex for super resolution (see Ref.\cite{DeAngelis2010}).}
\label{fig:4}
\end{figure}

In view of our aim of designing a plasmonic assisted IR near-field nanoprobe, we will now consider a rounded gold cone as displayed in 
Fig.~\ref{fig:4}a) with dimensions common to most IR scattering-based scanning near-field optical microscopy (IR s-SNOM) probes. In detail, the tip 
is 25~$\mu$m long with an full cone angle of 40$^\circ$ and a tip curvature radius of 40~nm. It has been shown that tips with an apex 
diameter of 80 nm can achieve a lateral optical resolution of about the same size\cite{Kopf2007}. 
Onto its surface we imprint our previously designed grating with dimensions \textit{D}~=~940~nm, \textit{a}~=~0.5 \textit{D}, and 
\textit{h}~=~\textit{D}. In contrast to our previous illustration of adiabatic field compression, here we keep the groove depth fixed 
and gradually reduce the outer tip radius instead (Fig. \ref{fig:4}b)). For this corrugated tip, we expect a higher field confinement 
but a field extent at the tip apex as big as with the uncorrugated version. In anticipation of this, we also consider a third structure, 
what we call a corrugated supertip (Fig. \ref{fig:4}c)), where we fit a 2.5~$\mu$m long sharp apex on top of the corrugated cone. This 
gold nanoantenna has a base diameter diameter of 300~nm, a full cone angle of 6.65$^\circ$, and a curvature radius of 5~nm. Such an 
acute structure can be fabricated by electron-beam induced deposition and has been utilized to achieve sub-10 nm resolution in the 
optical regime \cite{DeAngelis2010}. Due to the dimensions of our grating, it is not possible to corrugate the nanoantenna as well, so
the grating does not stretch all the surface up to the apex. 

\begin{figure}
\centering\includegraphics[width=\textwidth]{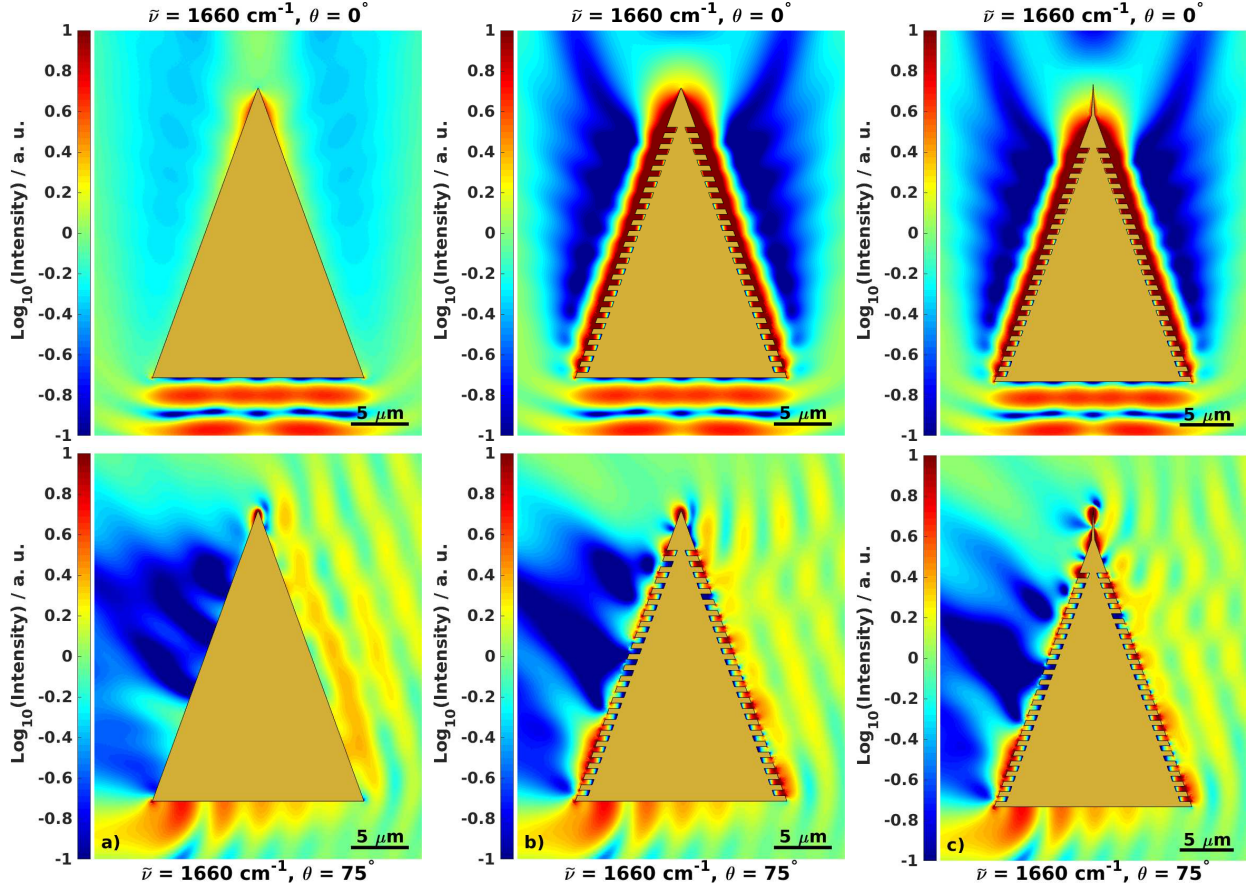}
\caption{FEM simulations of the electromagnetic field energy around the surface of a lossy a) uncorrugated gold tip, b) corrugated gold tip,
and c)corrugated supertip, all excited by p-polarized light at 1660 cm$^{-1}$ impinging from the bottom at an angle of 0$^{\circ}$ (top row) 
and 75$^{\circ}$ (bottom row) relative to tip axis.}
\label{fig:5}
\end{figure}

To determine how these three different structures perform in terms of confinement, we conducted FEM electromagnetic simulations 
where all tips were illuminated from the base by a p-polarized plane-wave with a frequency of 1660~cm$^{-1}$, first at angle of incidence of 
0$^{\circ}$ (Fig. \ref{fig:5}, top row) and then at 75$^{\circ}$ (Fig. \ref{fig:5}, bottom row) with respect to the tip axis, 
where the latter configuration has been used in all our previous experimental work 
\cite{Samson2006,Kopf2007,Wollny2008,Filimon2010,Ballout2011no2} as well by others \cite{Gomez2006, Kehr2011}. 
The field energy density distribution is normalized by the field energy density of the incident light. At first glance, 
there is no field enhancement at 0$^{\circ}$ incidence, even though we see for the corrugated structures excitation of spoof SPPs that 
are strongly confined to the grating. By contrast, we observe a field confinement at the tip apex for an angle of incidence of 75$^{\circ}$, 
which increases as we switch from the uncorrugated tip to the corrugated tip and then to the corrugated supertip. 

\begin{figure}
\centering\includegraphics[width=\textwidth]{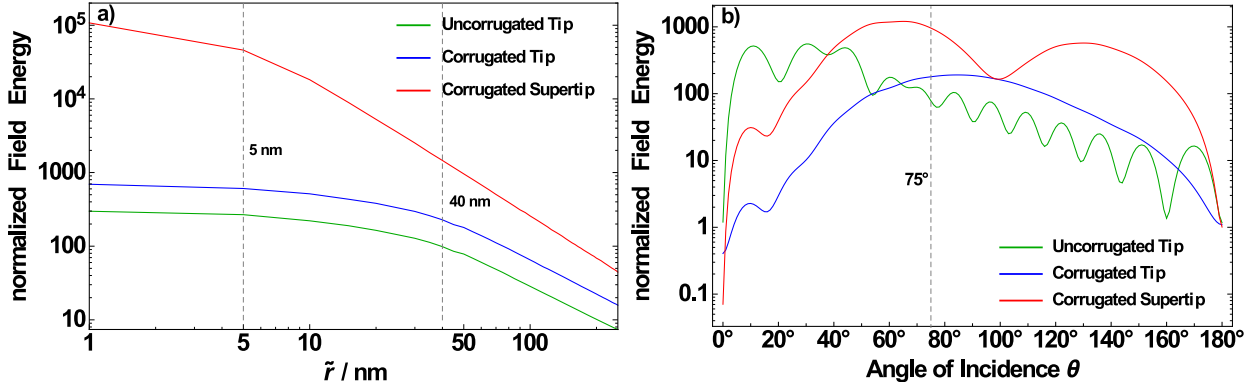}
\caption{a) Field energy as function of the radial distance from the tip apex in double logarithmic representation. The field confinement at the tip apex is measured as the 
extent at which the field drops down to $1/e$ of its value at the tip center (see dashed vertical lines). b) Field energy measured around 
the tip apex ($\tilde{r}=50$ nm) of all three tip configurations as function of the angle of incidence. The dashed vertical line 
indicates the angle of incidence usually used in our experimental setup.}
\label{fig:6}
\end{figure}

Before we eloborate on the angular dependence of the field enhancement, we first quantify the field confinement as an assessment for 
the achievable optical lateral resolution. As a measure we take the lateral field extent around the tip measured as the radial distance 
at which the field energy $\tilde{r}$ starts to follow a different power law visible as kinks in Fig. \ref{fig:6}a). Expecting the field extent 
to be comparable to the tip diameter we find the corrugated supertip to exhibit a sharp field distribution 
with a width of about 10 nm and a field enhancement of up to 3 orders magnitude larger compared to the ordinary tip, which provides a 
relatively poor lateral resolution of about 80 nm in agreement with the experimental findings mentioned earlier. Its corrugated counterpart 
provides the same lateral resolution, but about a factor of 2.5 better field enhancement thanks to the grating-induced SPPs. 

Now, regarding the angular dependence of the field enhancement as depicted in Fig. \ref{fig:6}b), the curve belonging to the uncorrugated gold tip 
features a multiple peaked, sloping distribution with the maximum at 30$^\circ$ similar to the study in Refs.~\cite{Krieger1990,Hagmann1997}. In 
agreement with antenna theory~\cite{Balanis2005}, the peaks of the multiple lobe radiation pattern of a long wire of 
length $l$ appear at angles of $\theta_m = \arccos\left[1\pm\left(2m+1\right)\lambda/\left(2l\right)\right]$ ($m=0,1,2,\dots$). Thus, the 
major radiation lobe ($m=0$) for a 25 $\mu$m long wire at an illumination wavelength of $\lambda = 6.024$ $\mu$m 
(i.e. $\tilde{\nu} = 1 660$ cm$^{-1}$) is expected to appear at 28$^\circ$. 

Up to the angle of grazing incidence (i.e. 20$^\circ$), the corrugated tip displays a similar angle dependence for the fields 
enhancement with the first peak around 15$^\circ$, but unlike the uncorrugated tip we notice at grazing incidence a rise of a shoulder 
rather than a local minimum. This can be explained by the fact that at grazing incidence the incoming p-polarized light has no field 
component along the surface, only along the surface normal. Thus, there is no field component to feed the antenna but only a component
to excite the SPP. Beyond grazing incidence the long-wire field pattern is overshadowed by a broad angular distribution with no distinct
features reflecting the almost omnidirectional excitation of spoof SPPs on the grating reaching a poorly pronounced/weak maximum at 
about 85$^\circ$, i.e. almost perpendicular to the wire axis. 

In contrast, the field enhancement around the tip apex of the corrugated supertip becomes more 
distinctive with peaks at an angle of incidence of 65$^\circ$ and 130$^\circ$ and a local minimum inbetween around 100$^\circ$. This is 
a result of the synergy of the antenna excitation and SPP excitation arising form the juxtaposition of an antenna and a grating in this 
single structure. From a geometrical analysis of the field components, we would expect 
the field enhancement around the apex of the corrugated supertip to peak at  65$^\circ$ and 155$^\circ$ with a 
local minimum in between at 110$^\circ$: Starting at grazing incidence where the electric field is entirely perpendicular to the 
surface, the ratio of the field components $E_{\parallel}/E_{\bot}=\tan\left(\theta-20^{\circ}\right)$ shifts in favour of 
$E_{\parallel}$ with increasing angle. At an angle of incidence of $\theta=65^{\circ}$ (i.e. 45 $^{\circ}$ 
relative to the tip surface) both field components, $E_{\parallel}$ and $E_{\bot}$, are even and thus both effects (antenna 
excitation and SPP excitation) should coherently contribute to the field confinement at the tip apex.
Beyond that angle, the ratio shifts even further in favour of $E_{\parallel}$ until the electric field vector is entirely parallel to 
the surface at 110$^{\circ}$ allowing only the antenna effect to contribute. As the angle increases further, the ratio now starts to shift 
in favour of $E_{\bot}$ and at 155$^{\circ}$ both field compenents are even again, thus maximizing again the cooperation of 
antenna and SPP excitation. 
However, our geometrical analysis does not entirely reproduce the observed angle distribution, particularly at obtuse angles, 
i.e. $\theta>90^{\circ}$. There, the characteristics of the antenna with its sloping distribution become more prominent making the 
peak at 155$^{\circ}$ not only to appear at about 15$^{\circ}$ smaller angle, but also to have a smaller height compared with the peak 
at 65$^{\circ}$. All in all, the angular distribution for the field enhancement of the corrugated supertip is a trade-off between the 
sharp dominant resonances of the long-wire antenna at acute and obtuse angles and the rather broad ``resonance'' of the grating at 
intermediate angles, reflecting the structural hybridisation of the antenna and the grating. 

\section{Conclusion}

By means of mode matching, we designed an experimentally feasible surface grating that excites spoof SPPs in the chemical
fingerprint region between 5.7 and 6 $\mu$m (i.e. between 1660 and 1755 cm$^{-1}$).
We grafted this grating onto the surface of a tapered metal waveguide structure to which we appended a nanoscale antenna with
a 5 nm tip radius. FEM electromagnetic simulations revealed extraordinary field enhancement of 2-3 orders higher compared with 
conventional IR near-field probes. This field enhancement is accompanied by field confinement at the tip apex of the order of 
10 nm holding the prospect of a fourfold increase in lateral optical resolution in comparison with conventional metal probes utilized in 
current s-SNIM apparatus. A goniometric analysis of the field excitation revealed a broad range of angles under which extraordinary 
field enhancement can be achieved. Thus, implementation of the nanoprobe presented here into existing near-field IR microscopy 
apparatus should be possible with minor or no modifications. With such a plasmon-assisted mid-IR nanoscopy probe available
the label-free imaging of single protein or lipid molecules within biological tissue is possible. Adapting the grating 
to the appropiate wavelength, such a probe can be also deployed to detect small fluctuations in the free carrier concentration of 
highly doped seminconductor nanostructures.

\section{Acknowledgement}
The authors thank R. P. Zaccaria (IIT Genova) and E. DiFabrizio (KAUST) for fruitful discussions and S. Burger (JCMWave GmbH) for 
the technical support with JCMSuite. The authors also thank the BMBF for the financial support under grant no. 05K10PCA. F. Ballout greatly 
appreciates the financial support under the ``Scheme to support special activities of doctoral students'' granted by the rectorate of 
the Ruhr-Universiy Bochum. M. Havenith acknowledges the financial support Single Molecule Detection (SMD), a project funded by 
the EU Commission under the 7th Framework Programme.

\end{document}